\documentclass[conference]{IEEEtran}
\pdfoutput=1
\IEEEoverridecommandlockouts
\usepackage{amsmath,amssymb}
\usepackage{graphicx}
\usepackage{booktabs}
\usepackage{multirow}
\usepackage{tikz}
\usepackage{xcolor}
\usepackage{url}
\usepackage{balance}
\usepackage{fancyhdr}
\usetikzlibrary{arrows.meta,positioning,fit,shapes.geometric,calc,backgrounds,decorations.pathreplacing}

\title{SentryBus: A Multi-Vantage Observability Model and Validated Instrument for I$^2$C Sensor-Interface Manipulation}

\author{
\IEEEauthorblockN{Sandesh More\IEEEauthorrefmark{1}, Elton Batista\IEEEauthorrefmark{1},Karla Daley\IEEEauthorrefmark{1}, Sneha Sudhakaran\IEEEauthorrefmark{1}}
\IEEEauthorblockA{\IEEEauthorrefmark{1}Department of Computer Engineering and Sciences\\
Florida Institute of Technology, Melbourne, FL, USA}
\thanks{Preprint notice. Manuscript submitted to the IEEE International Conference on Physical Assurance and Inspection of Electronics (PAINE) 2026. A revised version may differ from this preprint following peer review. Corresponding author: Sandesh More, smore2022@my.fit.edu.}
}

\begin{document}
\maketitle
\pagestyle{fancy}
\fancyhf{}
\fancyhead[L]{\footnotesize PREPRINT: SUBMITTED TO IEEE PAINE 2026}
\fancyhead[R]{\footnotesize \thepage}
\renewcommand{\headrulewidth}{0pt}
\fancypagestyle{plain}{\fancyhf{}\fancyhead[L]{\footnotesize PREPRINT: SUBMITTED TO IEEE PAINE 2026}\fancyhead[R]{\footnotesize \thepage}\renewcommand{\headrulewidth}{0pt}}
\thispagestyle{plain}

\begin{abstract}
Sensor-driven systems in medical Internet of Things devices, aerial drones, and cyber-physical systems commonly trust a measurement once it reaches the embedded processor. An adversary on the digital interface between a sensor and its processor can supply a plausible value that the correct firmware accepts and reports as ordinary telemetry. The hypothesis is that sensor interface manipulation leaves observable evidence at the acquisition path, that the evidence appears at different vantages depending on attacker position, and that a passive host-side monitor therefore has a measurable boundary beyond which manipulation becomes indistinguishable from legitimate acquisition. SentryBus models acquisition behavior on the I$^2$C sensor bus using transaction timing, read and write sequences, transfer lengths, address behavior, register and FIFO state access, and raw data transitions. The adversary is modeled as an inline interposer, parallel controller, sensor replacement, or compromised host because a commodity target-only sensor cannot initiate transfers or stretch, reorder, or delay bus transactions. A dual sided testbed captures both busses, host memory, and telemetry, and the detector consumes the host facing bus alone while remaining vantages serve as ground truth. A physiological instantiation reports three measured results: an inline interposer bounded at 0.842~percent of acquisition service time while preserving both the acquisition schedule and payload content, clean acquisition stability sustained over 6304~seconds at the telemetry vantage without a single clock regression, and a negative result establishing that data-transition features encode session specific signal statistics and do not transfer across capture sessions. Characterization of the capture instrument shows that a low-cost analyzer can truncate captures without kernel visible error and that vantage separation must be verified rather than assumed. Controlled attack trials are still outstanding, so no detection rate is claimed.
\end{abstract}


\section{Introduction}
A sensor value is often trusted because it arrives through the interface an embedded processor expects. A physiological module reports optical samples, a drone inertial unit reports acceleration, and an industrial sensor reports a process observation. Physical meanings differ, but the acquisition pattern is similar: a sensor is configured, sampled, and read over a peripheral interface before firmware converts raw bytes into a software-visible measurement.

The sensor-to-processor boundary creates a hardware security problem. When an adversary changes a transaction before the processor receives it, the firmware may process the altered value exactly as designed. A forged physiological sample can become IoMT telemetry, a modified inertial or barometric sample can influence a vehicle state estimate, and a replayed process value can reach a CPS monitoring or control application. In each case the application may remain uncompromised and the network message may be correctly formatted. Trust was lost earlier, at the hardware interface.

A commodity sensor is a passive I$^2$C target that does not initiate transfers and does not decide when it is read. The MAX30102 pin description lists SCL as an input only, so the device cannot hold the clock line low to stretch a transfer and cannot reorder or delay traffic on its own \cite{max30102}. Any autonomous buffering, reordering, delay, or replay therefore originates from an inline interposer or a compromised host, not from the sensor silicon.

The central observation is that evidence of a given manipulation does not appear uniformly. A response modified inside an inline interposer is visible on the host-facing bus. A configuration write altered before it reaches the sensor is visible only on the sensor-facing bus. A value changed inside host memory leaves no bus evidence at all. A synchronized replay may be indistinguishable at the bus and detectable only through host memory or application semantics. The research problem is therefore not merely whether attacks can be detected, but \emph{where the evidence of each attack first becomes observable and where a passive host-side monitor reaches its fundamental limit}.

SentryBus is a hardware-aware sensor-interface monitor and, more importantly, an instrument for a multi-vantage observability study. It models how a configured sensor communicates with its processor and treats a measurement path as suspicious when timing, address, operation sequence, transfer length, register or FIFO-state access, or raw-data transition behavior differs from known-clean acquisition. To keep the result honest, the detector observes only the host-facing bus, the vantage available to a deployable monitor placed beside the host.

The contributions are five. First, a multi-vantage acquisition model connecting sensor registers and FIFO, the sensor-facing bus, the host-facing bus, host memory, and telemetry, correcting a common misconception: a commodity target-only sensor cannot autonomously delay, reorder, or replay its own bus traffic. Second, a threat model organized by attacker electrical position with a seven-class attack taxonomy and stealth tiers. Third, a transparent host-side detector separating conformance violations from learned deviations, specified so inputs, thresholds, and training splits are auditable. Fourth, measured evidence that an inline interposer can be made transparent at the host vantage, bounding timing cost at 0.842~\% of service time while preserving schedule and payload content, alongside a measured negative result establishing that data-transition features do not transfer across sessions. Fifth, a characterization of the apparatus covering two silent failure modes, an analyzer truncating captures without kernel-visible error and a bypassed interposer yielding plausible single-vantage traces, with the checks detecting each.

\subsection{Scope and Status}
The paper delivers the threat model, the observability method, the measurement protocol, and a built and characterized instrument. Three empirical results are measured on the apparatus of Section~\ref{sec:testbed}: interposer transparency with bounded timing cost, long-run clean acquisition stability at telemetry, and a negative result on cross-session transferability of data-transition features. Controlled attack trials remain outstanding, so no detection rate and no observability vector is claimed, and Section~\ref{sec:futurework} states what each remaining result requires. Every hardware fact and prior-art claim is sourced to a datasheet, specification, or peer-reviewed publication.

\section{Background and Motivation}

\subsection{The Sensor-to-Telemetry Path and Four Observation Vantages}
\label{sec:vantages}
\begin{figure}[t]
\centering
\resizebox{\columnwidth}{!}{%
\begin{tikzpicture}[
font=\scriptsize,
box/.style={draw, rounded corners=2pt, minimum width=1.5cm, minimum height=.8cm, align=center},
vant/.style={draw, circle, fill=black!8, minimum size=.62cm, inner sep=0pt, font=\scriptsize\bfseries},
arr/.style={-{Latex[length=2mm]}, thick}
]
\node[box] (sen) {Sensor\\reg + FIFO};
\node[box, right=12mm of sen] (intp) {Inline\\interposer};
\node[box, right=12mm of intp] (mcu) {MCU +\\firmware};
\node[box, right=12mm of mcu] (net) {Telemetry};
\draw[arr] (sen)--node[above]{$V_S$}(intp);
\draw[arr] (intp)--node[above]{$V_H$}(mcu);
\draw[arr] (mcu)--(net);
\node[vant, below=7mm of sen] (vs) {$V_S$};
\node[vant, below=7mm of intp] (vh) {$V_H$};
\node[vant, below=7mm of mcu] (vm) {$V_M$};
\node[vant, below=7mm of net] (vt) {$V_T$};
\draw[dashed] (sen)--(vs);
\draw[dashed] (intp)--(vh);
\draw[dashed] (mcu)--(vm);
\draw[dashed] (net)--(vt);
\node[below=1mm of vh, font=\scriptsize\bfseries, align=center] {detector\\input};
\node[below=1mm of vs, font=\scriptsize, align=center] {ground\\truth};
\node[below=1mm of vm, font=\scriptsize, align=center] {ground\\truth};
\node[below=1mm of vt, font=\scriptsize, align=center] {ground\\truth};
\end{tikzpicture}}
\caption{Sensor acquisition path and four observation vantages. SentryBus computes detection features from the host-facing bus $V_H$ only. The sensor-facing bus $V_S$, host memory $V_M$, and telemetry $V_T$ are synchronized ground truth and are excluded from detector training and inference.}
\label{fig:vantages}
\end{figure}
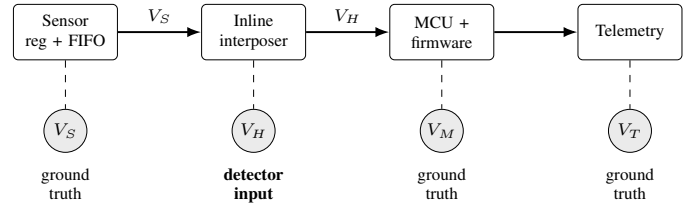

Connected sensing is often described as a data pipeline. At the device level it begins as a hardware acquisition path. Fig.~\ref{fig:vantages} shows the path and the four observation vantages used throughout the study. A physical phenomenon is sampled by a sensor that may perform conversion, filtering, or FIFO buffering. A processor retrieves registers or buffered samples over the bus, firmware converts the raw representation into a measurement, and an application packages the result for communication. An inline interposer separates the genuine sensor-facing bus $V_S$ from the host-facing bus $V_H$. Host memory $V_M$ and application telemetry $V_T$ complete the path.

Formally, for an attack $a$ the study records an observability vector
\begin{equation}
O(a)=\langle o_S,\,o_H,\,o_M,\,o_T\rangle,
\end{equation}
where each element indicates whether evidence is visible at that vantage and how early it appears. The vector, not a single accuracy number, is the primary result of the paper.

\subsection{FIFO Decoupling and Why Higher Layers Are Insufficient}
A FIFO decouples the time a sample was generated, the time the host retrieved it, and the time the application processed it. The MAX30102 stores up to 32 samples \cite{max30102} and the BMI160 provides a 1024-byte FIFO with hardware timestamps \cite{bmi160}. Decoupling is benign, so burstiness alone is not evidence of attack, and a detector must model expected periodicity together with FIFO-capacity constraints to separate malicious withholding or stale replay from ordinary buffering. Decoupling does not imply that a sensor pauses an active transaction and later releases traffic: clock stretching is optional in the specification \cite{i2cspec}, and the MAX30102 cannot stretch because SCL is an input only \cite{max30102}.

Sensor memory, bus state, and host memory form distinct layers. A commodity sensor exposes registers, FIFO buffers, interrupts, and status flags, not a stack or heap \cite{max30102,bmi160}; stack, heap, globals, and driver queues belong to the host runtime and become observable only through host instrumentation. Network-based anomaly detection suits attacks altering communication patterns \cite{said2021,aversano2023} but has little visibility into how a measurement was acquired, since telemetry produced from modified bytes still follows the expected schema, destination, and interval, and range checks share the limitation. Repeated acquisition creates a hardware behavioral structure narrower than the set of plausible values, so an attacker preserving the value format may still disturb timing, sequence, transfer length, or FIFO-state behavior.

\section{Related Work}

Network and application-layer anomaly detection for medical IoT observes above the peripheral interface: Said \emph{et al.} for smart hospital IoT \cite{said2021}, Aversano \emph{et al.} on medical IoT traffic \cite{aversano2023}, Sikder \emph{et al.} on device-level sensor context \cite{sixthsense}, and attention with physiological plausibility filtering for false-data injection \cite{dstanmed}. Sensor-network detectors model the temporal dynamics of reported readings, as in the first-order Markov transition framework of Mishra \emph{et al.} \cite{mishra2025wsn}, and consume values already delivered by the acquisition path. Each vantage sits above the acquisition path.

Spoofing and hardware-trust research establishes that software may trust a measurement whose sensing was already influenced: infusion-pump sensing manipulation \cite{park2016}, acoustic noise on gyroscopes \cite{son2015}, and acoustic injection into MEMS accelerometers \cite{walnut2017}, with transduction and noninvasive spoofing surveyed by Barua and Al Faruque \cite{sensorsurvey,barua2020}. Countermeasures include software detection of position-sensor spoofing \cite{tharayil2020}, active physical challenge-response \cite{pycra2015}, analog-Trojan detection on I$^2$C \cite{monjur2020}, IoT hardware-Trojan surveys \cite{sidhu2019}, spectral side-channel analysis \cite{syndrome2018}, and manufacturing-noise fingerprinting \cite{noisense2017}.

At the bus level, Cho and Shin fingerprinted electronic control units through clock-skew and message-interval timing \cite{cho2016}, motivating the timing and identity features used here. Work on MIL-STD-1553 spans security analysis \cite{milstd1553}, histogram-based intrusion detection \cite{maidens}, and AnoMili, combining physical intrusion detection, fingerprinting, and explanation \cite{anomili}. For the sensor interface specifically, Boone demonstrated TPM~Genie, an inline I$^2$C interposer intercepting and modifying host-to-target traffic \cite{tpmgenie}; Khelif \emph{et al.} defined non-invasive I$^2$C Trojan vectors \cite{khelif2021}; and Lorandel \emph{et al.} characterized wired interface traffic and hardware-attack detection \cite{lorandel2022}. Table~\ref{tab:relwork} positions SentryBus. The distinguishing combination is a sensor acquisition interface, an explicit attacker position, host-side detection with sensor-side ground truth, FIFO and register-state manipulation, and verified vantage separation.

Observation at the host memory vantage $V_M$ inherits the reliability limits of process memory acquisition. Tooling for recovering runtime objects from application memory bounds what such a vantage can yield \cite{ampledroid2020}, while Sudhakaran \emph{et al.} reported that recovered object state depends on acquisition conditions such as garbage collection and process state rather than on program semantics alone \cite{sudhakaran2022mem}, a dependence examined at length for Android application memory \cite{sudhakaran2022diss}. Fidelity at that vantage is further build dependent: symbol information required to interpret runtime structures has been progressively stripped from production platform binaries while the underlying memory architecture stayed stable, so an instrument calibrated on one build interprets another incorrectly \cite{nannapaneni2026}. Ali-Gombe \emph{et al.} paired memory-resident artifacts with learned representations for classification \cite{crgbmem2023}. The same caution governs the treatment of $V_M$ here: memory serves as instrumented ground truth inside a controlled testbed, not as a vantage a deployable monitor can rely on.

\begin{table}[t]
\caption{Positioning against the closest prior work}
\label{tab:relwork}
\centering
\scriptsize
\setlength{\tabcolsep}{3pt}
\begin{tabular}{lccccc}
\toprule
\textbf{Work} & \textbf{Iface} & \textbf{Interposer} & \textbf{FIFO/reg} & \textbf{Dual-side} & \textbf{Mem align} \\
\midrule
AnoMili \cite{anomili} & 1553 & no & no & partial & no \\
MAIDENS \cite{maidens} & 1553 & no & no & no & no \\
Lorandel \cite{lorandel2022} & I$^2$C & no & no & no & no \\
Khelif \cite{khelif2021} & I$^2$C & yes & no & no & no \\
TPM Genie \cite{tpmgenie} & I$^2$C & yes & no & no & no \\
Monjur \cite{monjur2020} & I$^2$C & yes & no & no & no \\
\textbf{SentryBus} & I$^2$C & yes & yes & yes & yes \\
\bottomrule
\end{tabular}
\end{table}

\subsection{Cross-Sensor Scope}
SentryBus considers physiological optical, inertial, and environmental sensing, treating IoMT, aerial, and CPS as motivating use cases rather than complete systems.

\begin{table}[t]
\caption{Sensor domains considered by SentryBus}
\label{tab:domains}
\centering
\footnotesize
\begin{tabular}{p{1.05cm}p{2.35cm}p{1.2cm}p{2.15cm}}
\toprule
\textbf{Domain} & \textbf{Sensor role} & \textbf{Bus} & \textbf{Acquisition behavior}\\
\midrule
IoMT & Optical physiological sensing & I$^2$C & FIFO and periodic register reads\\
Aerial & IMU and barometric sensing & I$^2$C/SPI & Burst and multi-axis reads\\
CPS & Process/environment sensing & I$^2$C/SPI & Periodic or event-driven reads\\
\bottomrule
\end{tabular}
\end{table}

The representation is shared, but the baseline is not. A high-rate IMU legitimately produces activity abnormal for a slow temperature sensor, so SentryBus learns robust centers, scales, and sequence frequencies per configuration. The cross-sensor claim forms a hierarchy: a common transaction representation is defensible, common feature families likely defensible, common attack signatures an empirical question, and universal transfer without re-baselining not claimed.

\section{Threat Model}
\label{sec:threat}
The adversary has physical or near-device access to a prototype, wearable, bedside, or edge sensing platform and can observe or interfere with the peripheral interface. Such access is consistent with reported practice on deployed consumer hardware rather than a laboratory abstraction: security assessment of consumer unmanned aerial platforms documents exposed onboard services and unauthenticated control paths \cite{more2025holystone}, earlier analysis of the same vendor family established practical data extraction from deployed units \cite{more2023thesis}, and multi-interface firmware acquisition on the same platform class recovers complete images through unprotected SPI flash and open debug ports without vendor cooperation \cite{more2026firmware}. Electrical position determines what is achievable, so five positions are distinguished. An inline interposer achieves response modification and substitution, replay, delay, withholding, and suppression. A command-modifying interposer achieves register-write modification, FIFO-pointer manipulation, and sample-rate change. A parallel controller achieves unauthorized read or write, collision, and bus occupation. A sensor replacement achieves same-address emulation, partial emulation, and removal. A compromised host achieves malicious polling, configuration change, and host-buffer modification. The command-modifying interposer is treated as a distinct experimental position because its manipulation appears on a different side of the bridge from response modification.

The attacks studied here are primarily inline-interposer attacks. Modifying a response while preserving address, length, and timing requires preventing the legitimate response from reaching the host, which a parallel device cannot reliably achieve because it contends electrically with the genuine sensor. The interposer presents a sensor-compatible interface to the host while communicating as a controller with the genuine sensor, and both sides are captured for ground truth. Detection features come exclusively from $V_H$. Attacks physically destroying the processor fall outside scope, and an adversary reproducing every electrical, temporal, structural, and data-dependent property of the legitimate sensor defines the upper bound of Section~\ref{sec:limitations}.

\section{Attack Taxonomy and Detectability}
Seven mechanisms are considered, each mapped to the attacker position realizing it and the vantage where evidence is expected first. \textbf{Injection} adds a forged transaction with data drawn from previously observed clean values, expected at $V_H$. \textbf{Modification} changes data bytes while preserving address and nominal length, expected at $V_H$ or $V_M$, and is the hardest case for structural features. \textbf{Replay} reproduces a captured sequence, and a synchronized interposer may align it with normal polling, so evidence is conditional or absent. \textbf{Timing and availability manipulation} delays, suppresses, or withholds traffic, expected at $V_H$; the interposer performs the delay, not the sensor \cite{max30102}. \textbf{Configuration and FIFO-state manipulation} alters mode, rate, FIFO control, or the read pointer, and moving the pointer backward causes stale samples to be reread \cite{max30102}; applied before the sensor, evidence appears at $V_S$ and may be invisible at $V_H$. \textbf{Removal} disconnects the sensor, expected at $V_H$. \textbf{Same-address substitution} replaces the sensor with a device answering at the same address and, if digitally faithful, may leave no evidence at any digital vantage.

A configuration write altered before the sensor illustrates conditional observability: the host-facing bus shows the legitimate command, the sensor-facing bus the malicious one, and a host-side detector sees only later behavioral consequences. The study therefore reports $P(\mathrm{detect}\mid\text{mechanism},\text{position},\text{monitor})$ rather than a single detection rate. Trials are stratified across four stealth tiers so reported detection is not dominated by obvious cases: T1 operationally obvious, T2 protocol preserving, T3 semantically plausible, T4 high-fidelity emulation. Payload strategy is separated from mechanism, ranging from random corruption through in-range values, low-and-slow drift, and physics-consistent values.

\section{SentryBus Design}

\subsection{Workflow and Detector Input}
Host-facing captures are decoded into transactions, grouped into acquisition windows, summarized by a feature extractor, and scored against a known-clean baseline. The sensor-facing bus, host memory, and telemetry enter only the ground-truth and validation path, never the detector.

\subsection{Representation, Features, and Scoring}
Decoding proceeds at three levels. A \emph{message} is a single address phase plus one directional byte sequence framed by START, optional repeated-START, and STOP; a \emph{register operation} combines a write-pointer message and a repeated-start read; an \emph{acquisition event} is one logical sampling operation containing several register operations or FIFO reads. Each message retains start and end times, address, direction, byte sequence, a per-byte acknowledgment vector, framing, and decoder errors. A final controller NACK after the last read byte is normal rather than an attack.

Six feature groups are extracted per window: temporal ($F_t$), operation sequence ($F_o$), transfer length ($F_l$), address ($F_a$), register and FIFO state ($F_r$), and data transition ($F_d$). The data-transition group summarizes change rather than value: for a raw field $x_i$ reconstructed from a consistent byte position, $\delta_i=x_i-x_{i-1}$ yields sign-change rate, repeated-run length, transition-magnitude quantiles, and short-window autocorrelation, exposing long exact replays without judging correctness.

A deterministic conformance layer encodes datasheet- and driver-derived expectations covering legal addresses, register sequences, transfer lengths, initialization order, FIFO states, and acknowledgment behavior, reporting violations directly. A statistical layer scores residual behavior with a per-feature distance appropriate to its type, using robust z-scores for scalar timing, quantile distance for timing distributions, Jensen-Shannon divergence for operation n-grams, and state-machine violation for FIFO transitions. Group scores convert to empirical percentiles from clean validation sessions and combine with fixed equal weights,
\begin{equation}
A(W)=\textstyle\sum_{g\in\{t,o,l,a,r,d\}} w_g\,E_g(W),\qquad \textstyle\sum_g w_g=1,
\label{eq:score}
\end{equation}
with a window flagged on a conformance violation or $A(W)>\tau$. The threshold $\tau$ is selected only from clean validation captures, never from attack data. The weighted score is not claimed as a contribution; the contribution is the observation model and the empirical findings.

\section{Experimental Testbed and Attack Realization}
\label{sec:testbed}
\begin{figure}[t]
\centering
\includegraphics[width=0.92\columnwidth]{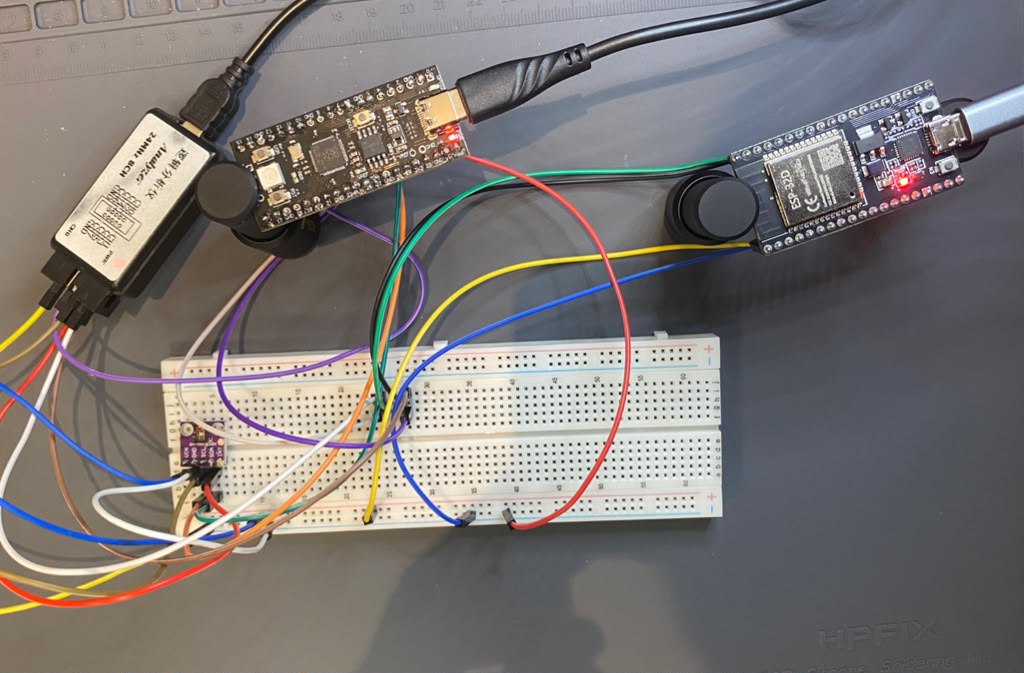}
\caption{Assembled physiological testbed. Left: fx2lafw logic analyzer providing common-clock dual-sided capture. Center-left on the breadboard: MAX30102 physiological sensor target. Center: RP2040 inline interposer presenting a sensor-compatible interface to the host while polling the genuine sensor as a controller. Right: ESP32 host running the fixed 50~Hz acquisition firmware. The interposer GPIO attack marker and both bus taps route to the analyzer.}
\label{fig:testbed_photo}
\end{figure}

The testbed design places one sensor per class behind an embedded host, with an inline interposer for attack realization and dual-sided passive probes on one multi-channel analyzer for a common clock. A GPIO marker asserted by the interposer at attack start is recorded by the analyzer to synchronize all artifacts. Six roles are built and measured: a MAX30102 target, an RP2040 hardware-I$^2$C interposer, an ESP32 host running fixed 50~Hz firmware, dual probes on one analyzer, a timestamped telemetry log providing $V_T$, and a sigrok and Python analysis chain. Three specified roles remain uninstantiated: a BMI160 inertial target, a non-FIFO control target, and the debug-capable host for $V_M$. Separating built from planned ensures no measurement is attributed to apparatus that does not exist. The primary interface is I$^2$C because the MAX30102 and BMI160 both support it and a common tuple with an address field maps cleanly onto I$^2$C; SPI is deferred because it lacks a standard address field and uses different framing. Fig.~\ref{fig:testbed_photo} shows the assembled physiological-pilot bench used for the measurements of Section~\ref{sec:testbed}.

\subsection{Three Interposer Modes and Bridge Overhead}
Because an interposer perturbs timing, the study runs three modes: direct sensor-to-host, transparent pass-through with the bridge present but benign, and attack. Bridge overhead is measured as
\begin{equation}
\Delta t_{\mathrm{bridge}}=t_{\mathrm{pass\text{-}through}}-t_{\mathrm{direct}},
\end{equation}
reporting its median, variance, and upper quantiles. Attack traces are compared primarily against transparent pass-through, because the physical bridge is present in both, so the detector cannot succeed merely by recognizing the bridge. To preserve realistic timing the interposer uses hardware I$^2$C or a deterministic state machine, begins at 100~kHz, and moves to a programmable-I/O or FPGA forwarder if synchronized 400~kHz forwarding cannot be maintained. The MAX30102 is a wearable-health module, not a certified diagnostic device, and is used only as a physiological testbed \cite{max30102}.

\subsubsection{Topology verification}
\label{sec:topology}
A dual-sided study is meaningful only when the probe pairs observe electrically distinct buses. An interposer powered on the bench with its sensor-facing pins reaching nothing would leave the host wired directly to the sensor, and both pairs would tap one node. Topology was therefore confirmed from the captures themselves by two independent tests.

The first test compares transaction rates across vantages within one capture. A five-channel bridge capture decodes to 497 acquisitions at 49.997~Hz on the host-facing bus and 1683 at 169.632~Hz on the sensor-facing bus. A shared node cannot carry two rates, so distinct rates within one file establish electrical separation directly. The second test removes the host: with the host disconnected, a capture of the sensor-facing bus alone still decodes 509 acquisitions addressed to the sensor at 169.632~Hz, traffic only the interposer acting as controller could generate.

The second test additionally confirms that the interposer sustains its own acquisition loop independently of host activity. Topology verification is reported because a bypassed interposer produces captures that decode without error and yield plausible timing statistics, a failure mode silent in exactly the way the capture truncation of Section~\ref{sec:instrument} is silent, so any reproduction should confirm vantage separation before interpreting a cross-vantage result.

\subsubsection{Measured bridge overhead}
A physiological pilot instantiates the testbed with a MAX30102 target, an RP2040 inline interposer, and an ESP32 host acquiring the six-byte FIFO payload at 50~Hz, captured at 8~MHz. Direct and pass-through captures at $V_H$ were compared per the definition above, taking as the compared quantity the service time of one acquisition, from the START of the FIFO pointer write to the STOP of the following read. Independent sessions share no time origin, so distributions were paired by order statistic at $n=497$ each. The direct baseline opens with a 50~ms startup transient across five acquisitions, excluded here.

\begin{table}[t]
\caption{Measured bridge overhead at vantage $V_H$, direct versus transparent pass-through}
\label{tab:bridgeoverhead}
\centering
\footnotesize
\begin{tabular}{lrr}
\toprule
\textbf{Quantity} & \textbf{Direct} & \textbf{Pass-through} \\
\midrule
Acquisitions scored & 497 & 497 \\
Median service time ($\mu$s) & 935.375 & 943.250 \\
Variance ($\mu$s$^2$) & 0.0025 & 3.852 \\
Standard deviation ($\mu$s) & 0.050 & 1.963 \\
$p_{90}$ / $p_{95}$ ($\mu$s) & 935.500 & 943.375 \\
$p_{99}$ ($\mu$s) & 935.500 & 943.500 \\
Acquisition period (ms) & 20.001 & 20.001 \\
Modal SCL clock (kHz) & 98.8 & 98.8 \\
\bottomrule
\end{tabular}
\end{table}

The rank-paired difference yields $\Delta t_{\mathrm{bridge}}$ with median 7.875~$\mu$s, variance 3.786~$\mu$s$^2$, and $p_{99}=8.000$~$\mu$s over a range of 3.500 to 8.125~$\mu$s, or 0.842~\% of direct service time. Two properties carry the transparency result. Upper quantiles sit within 0.25~$\mu$s of the median, so the interposer adds a near-constant offset rather than occasional stalls, and jitter rises from 0.050 to 1.963~$\mu$s while acquisition period and modal SCL clock stay fixed at 20.001~ms and 98.8~kHz. Pass-through therefore preserves the host-visible timing contract at the schedule level, perturbing it only within a transaction. Content transparency holds in parallel: reconstructed RED and IR means at both vantages agree to better than one count, 871.3 against 870.7 and 72.5 against 72.6.

The pilot interposer runs an acquisition loop asynchronous to the host, polling at 169.6~Hz against the host's 50~Hz and serving each host read from its most recent buffered sample rather than forwarding the transaction, so no store-and-forward latency exists to measure. The measurable quantity is data age, spanning 8 to 5877~$\mu$s with a median of 2933~$\mu$s and an upper bound matching the 5895~$\mu$s poll period. Data age is therefore bounded by one sensor-facing poll interval, a property relevant to the stale-sample and read-pointer mechanisms of Section~\ref{sec:threat}.

\section{Evaluation Protocol and Measured Results}
\label{sec:eval}

\subsection{Research Questions and Present Status}
\textbf{RQ1 (clean stability):} which interface properties remain stable across sessions, units, firmware executions, and configurations? \textbf{RQ2 (attack visibility):} which mechanisms and stealth levels are detectable from the host-facing bus? \textbf{RQ3 (cross-layer observability):} at which vantage does each attack first become distinguishable? \textbf{RQ4 (generalization):} which evidence categories generalize across sensors after calibration? The present study answers RQ1 in part, establishing a schedule stable to 50.00~Hz over 6304~s with zero clock regressions and demonstrating that $F_d$ does not transfer across sessions, while the remaining stability axes require multiple units and firmware configurations. RQ2 has its detector specified, thresholds fixed on clean data, and apparatus validated, but requires controlled attack trials stratified by tier. RQ3 has two bus vantages and telemetry instrumented and verified electrically distinct, and requires attack trials plus a debug-capable host for $V_M$. RQ4 states the claim hierarchy without claiming transfer, and requires inertial and non-FIFO targets. Stating status per question ensures that a claim the study does not support appears as an explicit absence rather than being obscured by adjacent supported claims.

\subsection{Protocol for Attack Trials}
For each attack the study aligns the marker, first bus evidence, host memory, and telemetry, measuring $L_{\mathrm{tele}}=t_{\mathrm{telemetry}}-t_{\mathrm{alert}}$, reporting a non-positive value honestly. Clean sessions span multiple units, power cycles, firmware executions, and bus clock rates, and hierarchical splitting tests on unseen sessions, units, and configurations, since random transaction-level splitting would exploit correlation between adjacent transactions. Planned comparisons place SentryBus against conformance-only, timing-only, transition-only, one-class, and n-gram detectors, with three negative controls: a host-memory-only modification leaving no bus evidence, a physical stimulus changing raw data while the bus stays normal, and a high-fidelity synchronized replay. Per Section~\ref{sec:validity}, every trial draws its baseline from its own session.

\subsection{Measured Clean Acquisition Stability at $V_T$}
A continuous telemetry log at $V_T$ characterizes host acquisition stability over an interval far longer than any bus capture. A run of 6304~s recorded 295079 lines, of which 295060 carried data, at a sustained 50.00 lines per second with zero monotonic-clock regressions and zero firmware restarts. The result supports RQ1 at telemetry: a correctly operating host holds its schedule over intervals three orders of magnitude longer than a detection window, so a schedule deviation at $V_H$ cannot be attributed to long-run drift.

\subsection{Capture Instrument Characterization}
\label{sec:instrument}
Reproduction depends on whether the capture instrument delivers the samples it reports, so the instrument was characterized rather than assumed correct. A low-cost FX2-class analyzer was exercised across a rate range spanning a factor of forty and a duration range spanning a factor of ten under two host software stacks, verifying every capture by comparing delivered sample count against requested duration times configured rate. Under favorable conditions the instrument delivered complete captures, and the measurements reported here were taken in that regime with counts matching exactly. Under degraded conditions it truncated every capture at a fixed wall-clock interval, near 0.5~s under one host stack and near 1.8~s under another, independent of configured rate across the full range. Independence from rate distinguishes a host-side transfer limit from an analyzer bandwidth ceiling, since the latter would truncate at a fixed byte count. Enumeration remained correct and no transfer error, stall, or reset appeared in kernel logs, so a truncated capture does not announce itself, and a delivered-count check must gate every capture.

\subsection{Ground-Truth Timing}
Attack start and end are recorded by a GPIO marker asserted by the interposer and sampled by the analyzer on a dedicated channel, so marker edges share the capture clock with both bus vantages and require no cross-machine synchronization. Direct measurement of the controlling host clock returned a monotonic tick of 15.6~ms, exceeding a full acquisition period at 50~Hz, so host-issued command timestamps serve only as a coarse cross-check and the marker channel remains the sole ground truth. Alignment of telemetry against bus evidence uses a deterministic content signature rather than a shared clock, because the manipulation under test flips a known bit position in a known payload byte.

\section{Discussion and Limitations}
An interface anomaly is evidence that acquisition differs from the clean baseline, not proof that a measurement is semantically false. The separation is a strength: a consistent hardware path with an unusual value warrants application-level handling, whereas an unusual value following an unexpected sequence, register write, or replay motif indicates acquisition-path interference. SentryBus serves as a laboratory assurance and forensic monitor for commodity sensors lacking authenticated measurements, complementing cryptographic authentication rather than replacing it.

\label{sec:limitations}
Baselines are sensor- and configuration-dependent, so a change in sampling frequency, FIFO mode, driver, or firmware may require re-baselining. A sophisticated interposer reproducing timing, structure, and electrical behavior may evade a host-side monitor; SentryBus increases the evidence an attacker must reproduce rather than providing authentication, and stronger assurance requires active challenge-response \cite{pycra2015} or clock and voltage fingerprinting \cite{cho2016}. Passive digital capture does not observe analog waveforms, and development sensors do not establish properties of certified devices.

\section{Threats to Validity}
\label{sec:validity}
Construct validity is addressed by comparing attack captures against transparent pass-through rather than direct mode, so detection cannot be credited to recognizing the interposer. Internal validity rests on instrument-derived ground truth: dual-sided capture, a hardware marker, and verified vantage separation. Attack labels remain authored decisions requiring preregistration or independent review. Guarding against a detector that learns one board requires evaluation across multiple units, sessions, and configurations with hierarchical splitting. External validity is bounded: development modules do not establish properties of certified systems, and cross-sensor transfer is a measured hypothesis. Decoder fidelity is a measurement threat, since a logic analyzer records digital edges rather than analog waveforms.

A further threat concerns baseline transferability and was surfaced by the physiological pilot. Scoring a transparent pass-through capture at $V_H$ against a clean baseline drawn from a different session returned a flag rate of 1.000 over 16 windows. Per-feature attribution located the entire deviation in the data-transition group $F_d$, with the timing and structural features $F_t$ and $F_r$ deviating by exactly zero, and the two vantages within the pass-through capture agreeing on reconstructed channel content to better than one count. The deviation therefore originates upstream of the interposer, at the sensor optical input, where the two sessions differed in illumination: consecutive-byte deltas were zero in 0.3~\% of the direct session and 28.3~\% of the pass-through session. The finding establishes that $F_d$ encodes session-specific signal statistics and does not transfer across capture sessions, so a data-transition score computed against a foreign baseline carries no detection meaning. Attack trials consequently require a clean baseline captured in the same session and under the same input conditions as the trial, and reporting the cross-session flag rate as evidence of interference would invert the result, since the transparency evidence resides in the features that deviate by zero.

\section{Reproducibility, Ethics, and AI Disclosure}
\label{sec:repro}
Upon publication the study releases raw captures, decoded traces, interposer and host firmware with debug symbols and linker map, analysis scripts, and a dataset card recording every provenance field. Each capture carries its delivered sample count and the vantage-separation evidence of Section~\ref{sec:topology}, so a third party can confirm instrument integrity and topology before interpreting any cross-vantage result. Numeric claims are accompanied by the clean-validation captures used to set thresholds.

The work uses commercial development modules on a laboratory bench, involves no patients or fielded platform, and uses a controlled optical fixture rather than human recordings. Attacks execute only on the authors' own hardware, and any previously unknown weakness found in a commercial sensor or driver goes to the vendor through coordinated disclosure. Generative AI tools assisted with drafting and literature triage; every factual and numerical claim was verified against a primary source, no result is reported that was not measured on the testbed described here, and the authors take full responsibility for all claims, citations, and conclusions.

\section{Future Work}
\label{sec:futurework}
The protocol is fully specified and the instrument characterized, so remaining work is execution rather than design. Modification, read-pointer manipulation, and removal at $V_S$ and $V_H$, with detector comparison and ablation, are blocked only by trial execution on apparatus already validated here. Host-buffer modification at $V_M$, same-address substitution, and cross-sensor transfer are blocked by hardware the bench lacks: a debug-capable host exposing a halted-core interface with symbols and linker map, a high-fidelity emulator, and inertial and non-FIFO targets. Pre-sensor configuration writes and synchronized replay require additional interposer firmware modes. An item blocked by execution rests on validated apparatus, whereas one blocked by hardware cannot be obtained here at any effort. Attack trials must draw the clean baseline from the same capture session, which the single-file protocol of clean, condition, and clean phases satisfies. Removing the hypervisor from the capture path is the next experiment for lifting the truncation ceiling of Section~\ref{sec:instrument}.

\section{Conclusion}
SentryBus reframes sensor-interface security as a multi-vantage observability study. A plausible sensor value should not inherit trust merely because correct firmware and telemetry process it. Measurements on a physiological instantiation establish that the inline instrument such a study requires can be made transparent, costing 0.842~percent of acquisition service time while leaving schedule and payload intact. A negative result carries equal weight: data-transition features encode session-specific statistics, so any trial using them needs a baseline from its own session, and a detector scored against a foreign session reports interference where none exists. Instrument characterization shows that a low-cost analyzer can truncate captures without kernel-visible error and that vantage separation must be verified rather than assumed, both failures being silent. The sensor-to-processor interface deserves explicit inspection as a hardware trust boundary.

\footnotesize
\balance

\end{document}